\documentclass{ws-procs10x7}
\usepackage{balance}
\usepackage{epsfig}
\usepackage{amssymb}
\usepackage{color}
\usepackage{url}
\usepackage[colorlinks]{hyperref}
\usepackage[italian,american]{babel}
\newcolumntype{d}[1]{D{.}{.}{#1}}

\def\ifm#1{\relax\ifmmode#1\else$#1$\fi}  

\def\x{\ifm{\times}}  
\def\pt#1,#2,{\ifm{#1\x10^{#2}}}

\def\ab{\ifm{\sim}}

\def\to{\ifm{\rightarrow}} 
\def\kl{\ifm{K_L}}   
\def\ks{\ifm{K_S}}
\def\kb{\ifm{\rlap{\kern.3em\raise1.9ex\hbox to.6em{\hrulefill}} K}}
\def\bN{\ifm{\rlap{\kern.3em\raise1.9ex\hbox to.6em{\hrulefill}} N}}
\def\ko{\ifm{K^0}}

\def\po{\ifm{\pi^0}}
\def\pic{\ifm{\pi^+\pi^-}}  
\def\K{\ifm{K}}
\def\rmk{\rm\kern.5mm }

\def\figbox#1;#2;{\parbox{#2cm}{\epsfig{file=#1.eps,width=#2cm}}}

\newcommand{\bra}[1]{\ensuremath{\langle\,#1\,|}}
\newcommand{\ket}[1]{\ensuremath{|\,#1\,\rangle}}

\newcommand{\Sec}[1]{Section~\ref{#1}}
\newcommand{\Secs}[1]{Sections~\ref{#1}}

\catcode`@=11 

\newdimen\z@ \z@=0pt 
\newskip\z@skip \z@skip=0pt plus0pt minus0pt
\def\m@th{\mathsurround=\z@}
\def\ialign{\everycr{}\tabskip\z@skip\halign} 
\def\eqalign#1{\null\,\vcenter{\openup\jot\m@th
  \ialign{\strut\hfil$\displaystyle{##}$&$\displaystyle{{}##}$\hfil
      \crcr#1\crcr}}\,}
\catcode`@=12 
\catcode`@=11 
\newcount\equanumber         \equanumber=0
\newdimen\referenceminspace  \referenceminspace=5pc
\let\cl=\centerline

\newcount\figurecount     \figurecount=0
\def\FIG#1{ \global\advance\figurecount by 1 \xdef#1{\the\figurecount}}
\newcounter{tablecntr}
\def\TAB#1{\stepcounter{tablecntr} \xdef#1{\kern.1ex\thetablecntr}}

\def\figboxc#1;#2;{\vglue2mm\cl{\figbox#1;#2;}\vglue3mm}
\def\captionfont{\small}
\let\cl=\centerline
\def\allcap#1;#2;{{\renewcommand{\baselinestretch}{.9}\captionfont
\newdimen\fcwidth \fcwidth=\textwidth \advance\fcwidth by -2cm
\setbox0=\hbox{{\bf Fig. #1.} #2}  
  \ifdim \wd0>\fcwidth  
       \vbox{\noindent
          \parshape=1 1truecm \fcwidth {\bf Fig. #1.} #2}
    \else
       \cl{{\bf Fig. #1.} #2}
    \fi}\renewcommand{\baselinestretch}{1.}\normalsize\hypertarget{#1}\vglue2mm }

\def\to{\ifm{\rightarrow}}    \def\plm{\ifm{\pm}}
\def\K{\ifm{K}}

\def\kpm{\ifm{K^\pm}}  

 \def\ab{\ifm{\sim}}  \def\x{\ifm{\times}}
  
\def\pt#1,#2,{\ifm{#1\x10^{#2}}}

\def\ord#1;{\ifm{{\mathcal O}(#1)}}

\renewcommand\section{\@startsection{section}{1}{\z@}%
                                    {-3.5ex \@plus -1ex \@minus -.2ex}%
                                    {2.3ex \@plus.2ex}%
                                    {\reset@font\large\bfseries\mathversion{bold}}}
\renewcommand\subsection{\@startsection{subsection}{2}{\z@}%
                                       {-3.25ex\@plus -1ex \@minus -.2ex}%
                                       {1.5ex \@plus .2ex}%
                                       {\reset@font\large\bfseries\mathversion{bold}}}
\renewcommand\subsubsection{\@startsection{subsubsection}{3}{\z@}%
                                          {.5ex \@plus .2ex}
                                          {-1.5em}
                                          {\reset@font\normalsize\sc}}

\renewcommand\paragraph{\@startsection{paragraph}{4}{\z@}%
                                          {.5ex \@plus .2ex}
                                    {-1em}%
                                    {\reset@font\normalsize\sc}}

\makeindex
\begin{document}

\title{$|V_{us}|$ measurement and CKM unitarity }

\author{M. Antonelli}

\address{Laboratori Nazionali di Frascati dell'INFN, Frascati, Italy.\\$^*$E-mail: Mario.Antonelli@lnf.infn.it}

\twocolumn[\maketitle\abstract{We review recent measurements of kaon decays and lifetimes 
 for the determination of the CKM element $V_{us}$. Their relevance to
 CKM matrix unitarity is discussed.}

]

\section{Introduction}
The realization that a precise test of CKM unitarity can be obtained 
from the first-row constraint
$|V_{ud}|^2 + |V_{us}|^2 + |V_{ub}|^2 = 1$ (with $|V_{ub}|^2$ negligible)
has sparked a new interest in good measurements of quantities related to
$|V_{us}|$.
As we discuss in the following sections, $|V_{us}|$ can be determined 
using semileptonic kaon decays;
the experimental inputs are the BRs, lifetimes, 
and form-factor slopes. One problem that frequently plagues the interpretation of older 
BR measurements is lack of clarity in the treatment of
radiative effects.
All new measurements of kaon decays with charged particles  are fully inclusive of radiation. 

\subsubsection{$K_{\ell3}$ decays}
\label{sec:vusl3}
The semileptonic kaon decay rates
still provide the best means for the measurement of $|V_{us}|$
because only the vector part of the weak current contributes
to the matrix element $\bra{\pi}J_\alpha\ket{K}$. In general,
\begin{displaymath}
\bra{\pi}J_\alpha\ket{K} = f_+(t)(P+p)_\alpha + f_-(t)(P-p)_\alpha,
\end{displaymath}
where $P$ and $p$ are the kaon and pion four-momenta, respectively,
and $t=(P-p)^2$.
The form factors $f_+$ and $f_-$ appear because pions and kaons are 
not point-like particles, and also reflect both $SU(2)$ and $SU(3)$ 
breaking. For vector
transitions, the Ademollo-Gatto theorem \cite{ag} ensures that
$SU(3)$ breaking appears only to second order in $m_s-m_{u,d}$. 
In particular, $f_+(0)$ differs from unity by only 2--4\%. 
When the squared matrix element is evaluated, a factor of $m_\ell^2/m_K^2$
multiplies all terms containing $f_-(t)$. This form factor can be 
neglected for $K_{e3}$ decays. For the description of $K_{\mu3}$ decays,
it is customary to use $f_+(t)$ and 
the scalar form factor $f_0(t) \equiv f_+(t) + [t/(m_K^2-m_\pi^2)]\,f_-(t)$. 

The semileptonic decay rates, fully inclusive of radiation, are given by
\begin{eqnarray}
\Gamma^i(K_{e3,\,\mu3})&=&|V_{us}|^2\:{C_i^2\:G^2\:M^5\over768\pi^3}\:S_{\rm
EW}\:|f^{\ko}_+(0)|^2 \nonumber \\
 & & I_{e3,\,\mu3}\: 
  (1+2\delta_{i,\,\rm em}+
2\delta_{i,\,SU(2)}).\nonumber
\end{eqnarray}
In the above expression, $i$ indexes $\ko\to\pi^\pm$ and $K^\pm\to\po$
transitions, for which $C_i^2 =1$ and 1/2, respectively. $G$ is the Fermi
constant, $M$ is the appropriate kaon mass, and $S_{\rm EW}$ is the
universal short-distance electro-week correction\cite{as}. 
The $\delta_{i,\,\rm em}$ terms are the long-distance
radiative corrections, which depend on the meson charges and lepton masses.
The form factors are written as 
$f_{+,\,0}(t)=f_+(0)\tilde f_{+,\,0}(t)$, with $\tilde f_{+,\,0}(0)=1$. 
$f_+(0)$ reflects $SU(3)$- and $SU(2)$-breaking corrections.
 $f_+(0)$ is calculated for $\ko$ transitions, 
 $\delta_{K^\pm,\,SU(2)}=(2.3\pm0.2)\%$\cite{aa} accounts for the 
 $SU(2)$-breaking corrections.
 $I_{e3,\,\mu3}$
is the integral of the Dalitz-plot density over the physical region
and includes $|\tilde f_{+,\,0}(t)|^2$.   
The vector form factor $f_+$ is dominated by the vector $K\pi$ resonances, the
closest being the $K^*(892)$.
The natural form for $\tilde f_+(t)$ is then:
\begin{equation}
\tilde f_+(t) ={M_V^2\over M_V^2-t}.\nonumber
\label{eq:pole}
\end{equation}
It is also customary to expand the form factor in powers of $t$ as
\begin{displaymath}
\tilde f_+(t)=1+\lambda'_+{t\over m^2_{\pi^+}}+
{\lambda''_+\over2}\left({t\over m^2_{\pi^+}}\right)^2.\nonumber
\end{displaymath}

\subsubsection{$K\to\mu\nu$ decays}
\label{sec:vusl2}
High-precision lattice quantum chromodymanics (QCD) results have 
recently become available and are
rapidly improving \cite{lat}. The availability of precise values for the
pion- and kaon-decay constants $f_\pi$ and $f_K$ allows use of a relation
between $\Gamma(K_{\mu2})/\Gamma(\pi_{\mu2})$ and $|V_{us}|^2/|V_{ud}|^2$,
with the advantage that lattice-scale uncertainties and radiative corrections
largely cancel out in the ratio \cite{ref:marfk}:
\begin{eqnarray}
{\Gamma(K_{\mu2(\gamma)})\over\Gamma(\pi_{\mu2(\gamma)})}&=&%
{|V_{us}|^2\over|V_{ud}|^2}\;{f_K\over f_\pi}\;%
{m_K\left(1-m^2_\mu/m^2_K\right)^2\over m_\pi\left(1-m^2_\mu/m^2_\pi\right)^2}\nonumber\\
& &\x(0.9930\pm0.0035),\nonumber
\label{eq:fkfp}
\end{eqnarray}
where the precision of the numerical factor due to structure-dependent 
corrections \cite{ref:fink} can be improved.

\subsection{$K_L$ decays}
\label{sec:kldec}
\subsubsection{$K_L$ lifetime}
\label{sec:KLlife}
KLOE has performed a fit to the proper-time distribution for 
$\kl\to3\po$ decays, which can be isolated with high purity and high and 
uniform efficiency.
The KLOE result is $\tau_{\kl}=50.92\pm0.30$~ns\cite{KLOE+05:KLlife}.
\subsubsection{$K_L$ BRs}
\label{sec:KLBR}
Recent measurements of dominant \kl\ branching ratios from KTeV, KLOE, and NA48
are in disagreement with avrages of previous measurements.

KLOE has performed a measurement of the absolute  \kl\ BRs 
\cite{KLOE+06:KLBR} of the four dominant(~99.5\% of all decays)
\kl\ decay modes. The presence
of a \kl\ is tagged by observation of a $\ks\to\pic$ decay. 
The value of these absolute BRs depend on the \kl\ lifetime value,
which enters into the calculation of the geometrical efficiency.
 This allows KLOE, by applying the constraint $\Sigma B=1$,
 to obtain another independent value of \kl\ lifetime:
$\tau_{\kl}=50.72\plm0.36$~ns.

A precise determination of the six largest \kl\ branching ratios has been obtaind by KTeV\cite{KTeV+04:BR}
by measuring the
following ratios: ${\rm B}_{\kl\mu 3}/{\rm B}_{\kl e3}$,
${\rm B}_{+-0}/{\rm B}_{\kl e3}$, 
${\rm B}_{000}/{\rm B}_{\kl e3}$,
 ${\rm B}_{+-}/{\rm B}_{\kl e3}$, and
${\rm B}_{00}/{\rm B}_{000}$.

 A measurement of the ratio B$_{Ke3}$/B$_{2T}$\cite{NA48+04:Ke3L} has been performed by NA48,
 where the 2T ``decay mode'' include all modes with 2 charged particles in the final
 state  B$_{2T}$=1.0048-B$_{000}$.
 A preliminary measurement of B$_{000}$ from NA48 is reported in Ref.\cite{NA48+04:ICHEP}.

 We perform an average of KLOE, KTeV, NA48 results, and the \kl\ lifetime
 measurement from Ref\cite{vos} by applying the 
 constraint that the \kl\ BRs must sum to unity. Correlation
 among channels are taken into account. The results
 are given in Tab.\ref{tab:kl}. 
\begin{table}
\tbl{Averages for \kl\ BRs(\%) and lifetime  \label{tab:kl}}
{\begin{tabular}{@{}lccccc@{}}
\toprule
  & B$_{\kl e3}$& B$_{\kl \mu 3}$ & B$_{000}$ & B$_{+-0}$&$\tau_L$(ns)\\ \colrule
 m & 40.46 & 26.97 & 19.69 & 12.52 &51.10\\
$\delta$ & 0.08 & 0.07 & 0.12 & 0.06&0.19\\
\botrule
\end{tabular}}
\end{table}

\subsection{$K_S\rightarrow \pi e\nu$}
\label{sec:kse3}
Using the \kl\ tag, obtained by observing
the interaction of a \kl\ in the calorimeter, KLOE has isolated a very pure sample of \ab13,000
semileptonic \ks\ decays and accurately measured the BR
for $\ks\to\pi e \nu$.
KLOE obtains ${\rm B}_{\ks e3} = (7.046\pm0.092)\times 10^{-4}$\cite{KLOE+06:KSe3}.

\subsection{$K^\pm$ decays}
\label{sec:kpm}
\subsubsection{$K^\pm$ lifetime}
The measurements of the \kpm\ lifetime listed in the
PDG compilation\cite{PDG06} exhibit poor consistency.
The KLOE preliminary result $\tau_\pm= 12.367\pm0.078$ ns,
 obtained from a fit to the proper time distribution, is in agreement with the
PDG average. 

\subsubsection{$K^{\pm}$ BRs}
A campaign for new measurements of the dominant charged kaon branching ratios
is ongoing at KLOE, NA48, and ISTRA+. In the following, we only consider measuremets
 in which the effects due to photon radiation have been explicitely 
 taken into account.

The ratio ${\rm B}_{K^\pm{e3}}/{\rm B}_{\pm 0}$ has been measured 
by E865 \cite{E865+03:Ke3}, NA48\cite{NA48+04:ICHEP}, and ISTRA+\cite{Kp:ISTRA}.

 Mesurements of ${\rm B}_{K^\pm{\mu3}}/{\rm B}_{K^\pm{e3}}$ have been
 performed by  NA48\cite{NA48+04:ICHEP}, and by E246\cite{E246}.

 KLOE has measured the absolute BRs
 of $K^+\to\mu \nu$\cite{KLOE+06:Kmu2}, $K^\pm\to \po e^\pm \nu$,
 $K^\pm\to \po \mu^\pm \nu$\cite{eps}, and  $K^\pm\to\pi^\pm \po \po$\cite{KLOE+04:taup}.

 We also consider the mesurement of ${\rm B}_{K^\pm{\mu 2}}/{\rm B}_{\pm 0}$
 inclusive of radiation performed by PS183\cite{PS183}, and the PDG average for
  $K^\pm\to\pi^\pm\pi^\pm\pi^\mp$\cite{PDG06}. 

 The average of the dominant $\K^\pm$ BRs
 is given in table in Tab.\ref{tab:kpmaverage}.
 We use the constraint that
 the $K^\pm$ BRs must sum to unity taking into account correlations.
  
\begin{table}
\tbl{Average for $K^\pm$ BRs(\%) and lifetime \label{tab:kpmaverage}}
{\begin{tabular}{@{}lccccc@{}}
\toprule
  & B$_{K^{\pm}e3}$& B$_{K^{\pm}\mu 3}$ & B$_{K^{\pm}\mu 2}$ & B$_{\pm 0}$&$\tau_\pm$(ns)\\ \colrule
 m       & 5.048 & 3.383 & 63.56 & 20.77 &12.384\\
$\delta$ & 0.031 & 0.027 & 0.13 & 0.12&0.022\\
\botrule
\end{tabular}}
\end{table}

\subsection{$K_{\ell3}$ form factor }
This measurement is particularly delicate.
Mainly because the effect of the form factor is maximal near the end of the $t$
 spectrum where the phase space goes to zero. 
\begin{figure}[h]
\centerline{\psfig{file=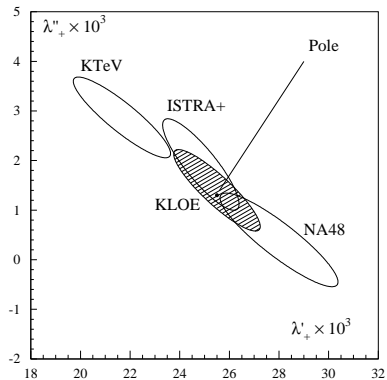,width=2.in}}
\caption{Experimental results for $\lambda_+'$ and $\lambda_+''$.}
\label{fig:ff}
\end{figure}

Figure \ref{fig:ff} compares KLOE results for 
$\lambda_+'$ and $\lambda_+''$ \cite{KLOE+06:KLe3FF}
with those from other experiments 
\cite{KTeV+04:FF,ISTRA+04:Kl3,NA48+04:Ke3FF}.
Measurements of the scalar form factor slope have been performed
by ISTRA+ using $K^\pm_{\mu 3}$, and KTeV using 
$K_{L\mu 3}$.
Since the experimental accuracies are larger than the
expected SU(2) breaking correction effects we average
measurements from  $K^\pm$ and \kl . The  result,
 with correlations taken into account, is given in Tab.\ref{tab:ff}.
\begin{table}
\tbl{Averages for form factor slopes \label{tab:ff}}
{\begin{tabular}{@{}ccc@{}}
\toprule
   $\lambda'_+$& $\lambda''_+$ & $\lambda_0$\\ \colrule
   $ 0.0250\pm0.0008$& $0.0016\pm0.0003$& $0.0158\pm0.0010$\\
\botrule
\end{tabular}}
\end{table}

\subsection{$V_{us}$ and CKM unitarity}
\label{sec:KLOEvus}
\begin{figure}[b]
\centerline{\psfig{file=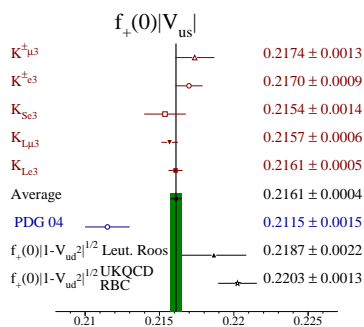,width=2.in}}
\caption{ $f^{\ko}_+(0)|V_{us}|$ for the five semileptonic
decay modes.}
\label{fig:f0vus}
\end{figure}

The available data set on $K_{\ell3}$ decays
allows multiple determinations of $|V_{us}|$.
Following the derivation in \Sec{sec:vusl3}, we compute the value
of $f^{\ko}_+(0)|V_{us}|$ from the decay rates for the five semileptonic
decay processes measured. The results are shown in Fig.\ref{fig:f0vus}.
We use the $K_S$ lifetime from the PDG \cite{PDG06}
while all other BR measurements and lifetime values are 
discussed in \Secs{sec:kldec},\ref{sec:kse3}, and \ref{sec:kpm}.

To extract the value of $|V_{us}|$, one needs an estimate of $f^{\ko}_+(0)$.
The original calculation of Leutwyler \& Roos \cite{leutR}, 
$f^{\ko}_+(0)=0.961\pm0.008$,  has been
confirmed recently by lattice QCD calculations
\cite{lattice_ab}. A considerable improvement, $f^{\ko}_+(0)=0.9680\pm0.0016$,
 has been achieved very recently \cite{lattice_UKQCD}.
 We obtain $|V_{us}|=0.2249\pm0.0019$ using $f^{\ko}_+(0)$ 
 from Ref.\cite{leutR}, or $|V_{us}|=0.2232\pm0.0006$ using 
 $f^{\ko}_+(0)$ from Ref.\cite{lattice_UKQCD}.
 To test CKM unitarity we use
 $\delta=|V_{ud}|^2+|V_{us}|^2+|V_{ub}|^2-1$, with
  $|V_{ud}|=0.97377\pm0.00027$ \cite{MS05}. We obtain
 a  good agreement with unitarity, $\delta = -0.0012\pm0.0010$,
 using $f^{\ko}_+(0)$ from  Leutwyler \& Roos \cite{leutR}.
  A ~3 $\sigma$ deviation is found, $\delta =-0.0020\pm0.0006$,
 using the recent calculation for $f^{\ko}_+(0)$ of 
 Ref.\cite{lattice_UKQCD}.

\begin{figure}[b]
\centerline{\psfig{file=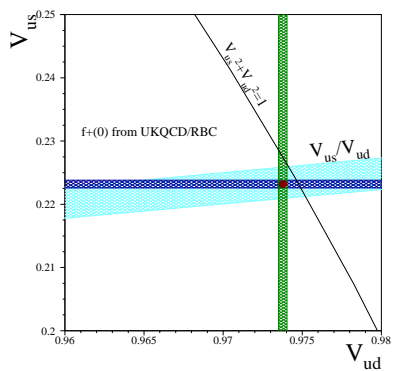,width=2.in}}
\caption{ Mesurements in the  $|V_{us}|$-$|V_{ud}|$ plane.}
\label{fig:vusvud}
\end{figure}

 Following the derivation in \Sec{sec:vusl2}, we  evaluate
 the ratio $|V_{us}|/|V_{ud}|$. We use the KLOE measurement of
 $K^+_{\mu2}$, all other experimental inputs from
 PDG, and  $f_K/f_\pi=1.1983^{+17}_{-7}$\cite{milc}.
  We obtain $|V_{us}|/|V_{ud}|=0.2294\pm0.0026$ in agreement
  with both values obtained from $K_{\ell3}$ decays.
 The interplay of all measurements is shown in Fig.\ref{fig:vusvud}.

\section*{Conclusions} 
 New precise exerimental results and lattice calculations
 allow to test of CKM unitarity at $10^{-3}-10^{-4}$ level. 
 A deviation of about 3$\sigma$,
 found using a new determination of  $f^{\ko}_+(0)$,
 has to be confirmed.

\end{document}